\documentclass[letterpaper, 10 pt, conference]{ieeeconf}  
\IEEEoverridecommandlockouts                              

\usepackage{graphics} 
\usepackage{epsfig} 
\usepackage{times} 
\usepackage{mathtools}
\usepackage{amsmath} 
\usepackage{amssymb}  
\usepackage{amsthm}
\usepackage{color}

\usepackage{mathrsfs}
\usepackage{csquotes}

\usepackage{tikz}
\usetikzlibrary{shapes,arrows}
\usepackage{pgfplots} 
\usepackage{pgfgantt}
\usepackage{pdflscape}
\usepackage{nicefrac}
\usepackage{prettyref}
\pgfplotsset{compat=1.5}
\pgfplotsset{plot coordinates/math parser=false}
\newlength\fwidth
\usepackage{grffile}
\usetikzlibrary{plotmarks}
\usetikzlibrary{arrows.meta}
\usepgfplotslibrary{patchplots}
\usepackage{units}
\let\labelindent\relax
\usepackage{enumitem}
\usepackage{hyperref}
\usepackage{float}
\usepackage{subcaption}
\usepackage{algorithm}
\usepackage{algpseudocode}
\usepackage{cite}

\newtheorem{remark}{Remark}
\newtheorem{theorem}{Theorem}
\newtheorem{lemma}{Lemma}
\newtheorem{problem}{Problem}

\newcommand{\T}{\scriptscriptstyle\top}       

\DeclareUnicodeCharacter{2212}{-}

\algnewcommand{\algorithmicgoto}{\textbf{go to}}%
\algnewcommand{\Goto}[1]{\algorithmicgoto~\ref{#1}}

\makeatletter
\def\widebreve{\mathpalette\wide@breve}
\def\wide@breve#1#2{\sbox\z@{$#1#2$}%
     \mathop{\vbox{\m@th\ialign{##\crcr
\kern0.18em\vspace{-0.016cm}\brevefill#1{0.55\wd\z@}\crcr\noalign{\nointerlineskip}%
                    $\hss#1#2\hss$\crcr}}}\limits}
\def\brevefill#1#2{$\m@th\sbox\tw@{$#1($}%
  \hss\resizebox{#2}{\wd\tw@}{\rotatebox[origin=c]{90}{\upshape(}}\hss$}
\makeatletter

\IEEEoverridecommandlockouts                              
\title{\LARGE \bf
High-Performance Trajectory Tracking MPC for Quadcopters with Coupled Time-Varying Constraints and Stability Proofs
}

\author{M. Izadi, A.T.J.R. Cobbenhagen, R.L. Sommer, {A.R.P. Andri\"en, E. Lefeber, W.P.M.H. Heemels}
\thanks{R.L. Sommer is with Avular Mobile Robotics, Eindhoven, The Netherlands 
(e-mail: {\tt\small r.sommer@avular.com}).}%
\thanks{The rest of the authors are with the Department of Mechanical Engineering, 
Eindhoven University of Technology, The Netherlands 
(e-mail: {\tt\small \{m.izadi.najafabadi1, a.t.j.r.cobbenhagen, 
a.a.j.lefeber, m.heemels\}@tue.nl}; 
{\tt\small a.andrien@gmail.com}).}%
}

\begin{document}

\maketitle
\thispagestyle{empty}
\pagestyle{empty}

\begin{abstract}
In this paper, we present a cascade control structure to address the trajectory tracking problem for quadcopters, ensuring uniform global asymptotic stability of the state tracking error dynamics. An MPC strategy based on a 12-dimensional prediction model is proposed for the outer loop, explicitly accounting for time-varying \enquote{coupled} constraints, where multiple variables are interdependent and need to be handled together. The outer-loop controller generates an acceleration reference, which is then converted into attitude and angular velocity references, later tracked by a nonlinear inner-loop controller. Numerical simulations validate the approach, demonstrating enhanced performance in precise and fast tracking by imposing less conservative constraints than existing approaches, while still guaranteeing stability.
\end{abstract}

\section{INTRUDUCTION}
\label{sec:sec1}
Quadcopters are among the most widely used Unmanned Aerial Vehicles (UAVs) due to their versatility in performing a range of tasks, including search and rescue operations, agriculture, civil, military missions, and medicine delivery. This growing interest has driven research on advanced control algorithms that deliver optimal performance in terms of fast response, stability, and precise tracking.

Trajectory tracking control of the quadcopter is challenging due to its underactuated configuration, along with nonlinear and unstable dynamics. The control problem of the quadcopter has been widely
investigated using various control approaches, including PID \cite{survey}, model predictive control (MPC) \cite{feedbacklinearization}, iterative learning control \cite{survey}, nonlinear control \cite{ErijenNLcontroller}, and sliding mode control \cite{survey}. Among these methods, 
MPC stands out for meeting four key features of a desirable controller: (i) anticipating future reference information, (ii) handling state and input constraints, (iii) real-time implementability on embedded hardware, and (iv) the availability of theoretical frameworks for stability guarantees.

Recent studies have made notable progress in both linear and nonlinear MPC for quadcopter control. In \cite{sadi2024cascade}, a hierarchical cascaded linear MPC controller is proposed for a quadcopter, dividing the overall MPC strategy into translational and rotational schemess, reducing computational complexity. In \cite{feedbacklinearization}, feedback linearization is applied to transform the quadcopter's nonlinear model into a linear system, forming the basis for linear MPC controller design. 
However, the linear MPC schemes proposed in both cited papers lack formal stability guarantees. \cite{NMPC_1} proposes a stability-guaranteed Nonlinear Model Predictive Control (NMPC) scheme for quadcopter trajectory tracking, introducing a distinct running cost to ensure stability. In \cite{NMPC_2}, a stability constraint ensures closed-loop stability for the NMPC-based quadcopter control. However, both NMPC approaches lack a formal stability guarantee due to using forward Euler discretization instead of an exact method.

A cascade control approach, recently proposed in \cite{Alexpaper}, employs MPC for outer-loop position control and a nonlinear controller for inner-loop attitude control. This method ensures uniform almost global asymptotic stability (UaGAS) while handling thrust constraints. However, its real-world performance is compromised due to conservatism introduced by the addition of an artificial stabilizing constraint and the transformation of nonlinear model constraints to linear ones. Building on the method proposed in \cite{Alexpaper}, \cite{izadi2024} presents a significantly improved MPC-based controller design for the outer loop, achieving higher performance. This substantial improvement is achieved by imposing time-varying (TV) decoupled constraints instead of more conservative time-invariant ones and by leveraging the marginal stability \footnote{All eigenvalues lie within the unit circle and those on the unit circle are simple.} property of the outer-loop dynamics to specify an appropriate terminal cost, rather than introducing conservatism through stabilizing constraints. \cite{izadi2024} decouples both dynamics and constraints to reduce online computational effort and enhance real-time implementability, resulting in three separate MPC formulations for the $x$, $y$, and $z$ directions with TV decoupled constraints. Despite significant improvement, the conservatism of decoupled constraints still limits performance.

Motivated by this gap, we propose a novel MPC scheme for outer-loop control with TV linear coupled constraints. The translational system, is modeled as a 12th-order linear system, with the non-zero total thrust constraint from the original nonlinear model translated into TV linear coupled constraints. An MPC controller is employed to generate a twice-differentiable virtual acceleration, which is then applied to compute the thrust control input. The desired virtual acceleration is converted into the desired attitude, which is tracked by the nonlinear attitude controller from \cite{ErijenNLcontroller} in the inner loop. Similar to \cite{izadi2024}, we leverage the controllability and marginal stability of the outer-loop dynamics, utilizing a specific cost function defined in \cite{neutrallystable} to ensure uniform global asymptotic stability (UGAS) of the outer loop. Compared to \cite{izadi2024}, we propose a more efficient outer-loop MPC strategy by introducing less conservatism through the use of TV \textit{coupled} constraints, rather than TV \textit{decoupled} ones, while still ensuring UGAS and inter-sample constraint satisfaction.
\section{PRELIMINARIES}
This section introduces the notation and theorems used in the paper. $\mathbb{R}_{>0}$, and $\mathbb{R}_{\geq 0}$ denote the set of positive real numbers, and the set of non-negative numbers, respectively. The set of natural numbers is denoted by $\mathbb{N} = \{1,2,3 \hdots\}$, and $\mathbb{N}_0 = \mathbb{N} \cup \{0\}$. 
This work considers UGAS, 
and uniform local exponential stability (ULES), with their definitions provided in \cite{Khalil:2002}, and adopts UaGAS, which is UGAS except for a measure-zero set of initial conditions, as defined in \cite{ErijenNLcontroller}. The saturation function $\mathrm{sat}:\mathbb{R}^m\to\mathbb{R}^m$ is defined as $ \mathrm{sat}(u)= [\mathrm{sat}(u_1),~ \mathrm{sat}(u_2),\hdots,~ \mathrm{sat}(u_m)]^{\T},$
where
\begin{equation}
\label{sat function}
{\mathrm{sat}(u_i)} = \begin{cases}
u_{\mathrm{max}},&\ u_i>u_{\mathrm{max}}\\ 
u_i,&\ |u_i|\leq u_{\mathrm{max}} \\
-u_{\mathrm{max}},&\ u_i<-u_{\mathrm{max}} 
\end{cases}
\end{equation}
for $i\in\{1,2,...,m\}$, where $u_{\mathrm{max}}>0$. Consider the discrete-time system
\vspace{-0.2 cm}
\begin{equation}
    x^+=Ax+B\mathrm{sat}(u),
    \label{saturated system}
\end{equation}
where $x\in\mathbb{R}^n$ and $u\in\mathbb{R}^m$ denote the  current states and inputs, respectively. It is assumed that $(A,B)$ is controllable, and the system is marginally stable. In this paper, the stability proof of the proposed MPC strategy is built on the theorems from \cite{neutrallystable}, one of which is repeated here for self-containment.

Marginal stability implies the existence of a positive definite matrix $M_c$, such that
\begin{equation}
    A^{\T} M_c A - M_c \preceq 0.
    \label{Mc}
\end{equation}
A globally stabilizing small-gain control is defined as
\begin{equation}
    u(x)=-\kappa B^{\T} M_c A x =: Kx,
    \label{smallgain}
\end{equation}
where $\kappa>0$ is chosen such that
\begin{equation}
    \kappa B^{\T} M_c B \prec I.
    \label{kappa}
\end{equation}
There exists a positive definite matrix $M_q$ such that (\hspace{-0.001cm}\cite{neutrallystable})
\begin{equation}
    (A+BK)^{\T} M_q (A+BK) - M_q = -I.
    \label{Mq}
\end{equation}
\begin{theorem}
(\hspace{-0.001cm}\cite{neutrallystable}).
    For the closed-loop system \eqref{saturated system} with the small-gain control \eqref{smallgain}, assuming that $(A,B)$ is controllable and \eqref{saturated system} is marginally stable, there exist a global Lyapunov function $W: \mathbb{R}^n \rightarrow \mathbb{R}_{>0}$ such that 
\begin{align}
&W(x)=W_q(x)+\lambda W_c(x)= x^{\T} M_q x+\lambda(x^{\T} M_c x)^{3/2} \label{LyapFunc}\\
&W(Ax+B\mathrm{sat}(Kx))-W(x)\leq -\|x\|^2, 
\end{align}
with K in \eqref{smallgain}, $M_c \succ0$ as in (\ref{Mc}), $M_q\succ 0$ as in (\ref{Mq}), and 
    \begin{equation}
        \lambda=\frac{2 \kappa L_u \sigma_{\mathrm{max}}(A_d^{\T} M_q B_d)}{\sqrt{\lambda_{\mathrm{min}}(M_c)}},
        \label{lambda eq}
    \end{equation}
    where $\sigma_{\mathrm{max}}$ and $\lambda_{\mathrm{min}}$ denote the maximum singular value and minimum eigenvalue of a matrix, respectively and $L_u$ is chosen such that $L_u u_{\mathrm{max}}>1$.
\label{Thoery small gain}
\end{theorem}

\section{DYNAMICS AND PROBLEM
FORMULATION} \label{sec:sec2}
\vspace{-0.1cm}
The model used in this work is based on the one presented in \cite{Alexpaper} and \cite{izadi2024}. Let $G$ be a right-handed inertial (or world) frame using the North-East-Down convention, with $\{x_G, y_G, z_G\}$ as its orthonormal basis vectors. $B$ represents a right-handed body-fixed frame with orthonormal basis vectors $\{x_B, y_B, z_B\}$, representing the axes of $B$ relative to $G$. The rotation matrix  $R = [x_B, y_B, z_B] \in SO(3)$ defines the orientation of the body-fixed frame relative to the world frame. Angular velocity, linear velocity and position of the body with respect to the world frame are described with vectors $\omega = [\omega_1, \omega_2, \omega_3]^{\T}$, $v = [v_x, v_y, v_z]^{\T}$ and $p = [p_x, p_y, p_z]^{\T} \in \mathbb{R}^3$, respectively.
With the defined variables, the dynamics of the quadcopter can be described as
\vspace{-0.15 cm}
\begin{subequations}\label{dynamics}
\begin{align}
 \Dot{p}&=v, \label{position}\\
\Dot{v}&= g z_G-T z_B - Dv, \label{velocity}\\
\Dot{R}&=RS(\omega), \label{Rotation}\\
J\Dot{\omega}&=S(J\omega)\omega-\tau_g-AR^{\T}v-C\omega+\tau, \label{angularvalocity}
\end{align}
\end{subequations}
where $g$ is the gravitational acceleration, $T \geq 0$ is the magnitude of the combined thrust of the four propellers, normalized by mass, $D = \operatorname{diag}(d_x, d_y, d_z)$, $d_x, d_y, d_z> 0$, are the mass-normalized rotor drag coefficients, $\tau_g \in \mathbb{R}^3$ are torques resulting from gyroscopic effects, $J \in \mathbb{R}^{3 \times 3}$ is the inertia matrix, $A$ and $C$ are constant matrices, \mbox{$\tau = [\tau_1, \tau_2, \tau_3]^{\T} \in \mathbb{R}^3 $} is the torque input and $S(\omega)$ represents a skew-symmetric matrix such that $S(a)b=a \times b$ for any vectors $a,b \in \mathbb{R}^3$.

The thrust is non-negative and limited according to
\begin{equation}
    0 \leq T(t) \leq T_{\mathrm{max}},~~ \text{for~all~} t \in \mathbb{R}_{\geq 0},
    \label{constraint}
\end{equation}
where $T_{\mathrm{max}} > g$. This physical restriction arises because the propellers can generate only a limited upward thrust and must be able to counteract the gravitational force.

A reference trajectory $\big(\bar{p}, \bar{v}, \bar{R}, \bar{\omega}, \bar{T}(t), \bar{\tau}\big) : \mathbb{R}_{\geq 0} \rightarrow \mathbb{R}^3 \times \mathbb{R}^3 \times SO(3) \times \mathbb{R}^3 \times \mathbb{R} \times \mathbb{R}^3 $ is referred to be a feasible trajectory, if it satisfies (\ref{dynamics}) for all $t \in \mathbb{R}_{\geq 0}$, i.e.,
\begin{subequations}\label{reference}
\begin{align}
\Dot{\bar{p}} &= \bar{v}, 
    & \Dot{\bar{v}} &= g z_G - \bar{T} \bar{z}_B - D\bar{v}, \label{pv ref} \\
\Dot{\bar{R}} &= \bar{R}S(\bar{\omega}), 
    & J\Dot{\bar{\omega}} &= S(J\bar{\omega})\bar{\omega} - \tau_g - A\bar{R}^{\T}\bar{v} - C\bar{\omega} + \bar{\tau}, \label{RW ref}
\end{align}

\end{subequations}
and $0 < \epsilon_1 \leq \bar{T}(t) \leq T_{\mathrm{max}}-\epsilon_2,$
with fixed $\epsilon_1, \epsilon_2>0$.

For a feasible reference trajectory, the error coordinates are defined as
\begin{equation}
 \label{error dynamics}
 \Tilde{p}=\bar{p}-p,~~ \Tilde{v}=\bar{v}-v,~~\Tilde{R}=\bar{R}^{\T} R,~~
 \Tilde{\omega}= \omega-\Tilde{R}^{\T} \bar{\omega}.
\end{equation}
The main tracking control problem of the quadcopter is:
\begin{problem}
\label{problem1}
    (\hspace{-0.0003 cm}\cite{Alexpaper}). Given a feasible reference trajectory $(\bar{p}, \bar{v}, \bar{R}, \bar{\omega}, \bar{T}, \bar{\tau})$, develop appropriate control laws
\begin{align*}
    T=T(p,v, R, \omega, \bar{p}, \bar{v}, \bar{R}, \bar{\omega}),~~
    \tau=\tau(p,v, R, \omega, \bar{p}, \bar{v}, \bar{R}, \bar{\omega}),
\end{align*}
such that (\ref{constraint}) holds and  the origin $(\Tilde{p}, \Tilde{v}, \Tilde{R}, \Tilde{\omega})=(0, 0, I, 0)$ of the resulting closed-loop system is UaGAS.
\end{problem}

\section{METHODOLOGY} \label{sec:sec4}
To solve Problem \ref{problem1}, a cascade control architecture is employed, consisting of an outer-loop and an inner-loop problem, that contain the translational \eqref{position}, \eqref{velocity} and rotational \eqref{Rotation}, \eqref{angularvalocity} dynamics, respectively. This section outlines the cascade configuration and presents the problem definitions for both the inner-loop and outer-loop  problems.
\subsection{Cascade trajectory tracking setup and constraints}
\label{section 4.1}

The dynamics for position and velocity errors in (\ref{error dynamics}) follow from  (\ref{pv ref}), (\ref{position}), and (\ref{velocity}), are given by 
\begin{subequations}
\begin{align*}
\Dot{\Tilde{p}}=\Tilde{v},~
\Dot{\Tilde{v}}=-D\Tilde{v}+Tz_B-\bar{T}\bar{z}_B.
\end{align*}
\end{subequations}
A desired acceleration error $a_d \in \mathbb{R}^3$ is defined as
\begin{equation}
    a_d=Tz_B-\bar{T}\bar{z}_B.
    \label{ad virtual}
\end{equation}
So, $T$ and $z_B$ have to be chosen such that \eqref{ad virtual} holds. This allows the translational system, including the position and velocity dynamics, to be reformulated as a linear system
\begin{equation}
   \label{outerloop dynamic} 
\Dot{\Tilde{p}}=\Tilde{v},~
\Dot{\Tilde{v}}=-D\Tilde{v}+a_d.  
\end{equation}

Thus, the first input $T$ can be determined by
\begin{equation} \label{thrust}
    T=\|a_d+\bar{T}\bar{z}_B\|.
\end{equation}
To track the desired virtual acceleration generated by the MPC in the outer loop, a desired thrust vector is generated and converted into a desired attitude $R_d$ and a desired angular velocity $\omega_d$, with $T$ applied to the quadcopter. The inner-loop attitude controller uses $R_d$, $\omega_d$, and measured attitudes and angular velocities to generate torque inputs $\tau$ for the quadcopter. In \cite{Alexpaper} and \cite{izadi2024}, it is shown how $a_d$ can be converted to the desired attitude $R_d$. Since the inner-loop controller requires setpoints for both the desired angular velocity and its derivative, the desired acceleration must be twice differentiable. For more details, refer to \cite{Alexpaper} and \cite{izadi2024}.

The constraint (\ref{constraint}) can be converted into a constraint on $a_d$ by considering the following set \cite{Alexpaper}:
\begin{multline}
\label{main ad constraint}
\mathcal{A}(\bar{R}, \bar{T})=\{a_d \in \mathbb{R}^3 ~|~ 0 <  \| a_d+\bar{T} \bar{z}_B  \| \leq T_{\mathrm{max}},\\
\bar{z}_B^{\T} a_d+\bar{T}>0 \}.
\end{multline}

\subsection{Cascade Problem Definition}

To address Problem \ref{problem1} using a cascade approach, we define two sub-problems: The outer-loop and inner-loop problems.
\begin{problem}[Outer-loop problem]
\label{outerloop problem}
Find a virtual acceleration control law $a_d = a_d(p,v,\bar{p}, \bar{v}, \bar{T})$, which is twice differentiable with respect to time, such that the origin $(\Tilde{p}, \Tilde{v})=(0,0)$ of the system (\ref{outerloop dynamic}) is UGAS and $a_d(t) \in \mathcal{A}(\bar{R}, \bar{T})$ for all $t \in \mathbb{R}_{\geq 0}$, with ${A}(\bar{R}, \bar{T})$ as defined in \eqref{main ad constraint}.
\end{problem}
Given the goal of steering $\Tilde{R}$ to $R_d$ in the inner loop, attitude and angular velocity errors are formulated as 
\begin{subequations}\label{Re-we}
\begin{align}
R_e&= R_d ^{\T} \Tilde{R}, \label{Re} \\
\omega_e&=\omega-\Tilde{R}^{\T} \bar{\omega}-R_e^{\T} \omega_d. \label{we} 
\end{align}
\end{subequations}
The inner-loop problem is now defined as follows:
\begin{problem}[Inner-loop problem]
\label{innerloop problem}
Find a control law $\tau = \tau(R,\omega,\bar{R}, \bar{\omega}, \bar{T}, a_d)$, such that the origin $(R_e,\omega_e)=(I,0)$ of the system (\ref{Re-we}) is UaGAS.
\end{problem}
\section{INNER-LOOP TRACKING}
\label{inner loop}
In this work, the non-linear controller from \cite{ErijenNLcontroller} is employed for stabilizing the attitude dynamics of the quadcopter, as it ensures ULES and UaGAS for the attitude dynamics.
The input is given by
\begin{align}
\label{tau}
\begin{split}
\tau&=-K_{\omega}\omega_e + K_R \sum_{i=1}^{3} k_i(e_i \times R_e^{\T} e_i)\\
&-S(J\omega)\omega + \tau_g +AR^{\T} v +C \omega \\
&+ J\Tilde{R}^{\T} J^{-1}(S(J\bar{\omega})\bar{\omega}-\tau_g - A\bar{R}^{\T} \bar{v} - C \bar{\omega}+\bar{\tau}) \\
&-J[(S(\omega)\Tilde{R}^{\T}-\Tilde{R}^{\T} S(\bar{\omega}))\bar{\omega}+S(\omega_e)R_e^{\T}\omega_d-R_e^{\T} \Dot{\omega_d}],
\end{split}
\end{align}
which results in the closed-loop system
\begin{subequations}
\begin{align*}
\Dot{R_e}=R_e S(\omega_e),~~
J\Dot{\omega}_e= -K_{\omega}\omega_e + K_R \sum_{i=1}^{3} k_i(e_i \times R_e^{\T} e_i), 
\end{align*}
\end{subequations}
for some $k_i>0$, $K_{\omega}>0$ and $K_R>0$. Since \( R_e \) converges to \( I \) for almost all initial conditions, it can be concluded from (\ref{Re}) that \( \Tilde{R} \) converges to \( R_d \) for almost all initial conditions. Additionally, as \( \omega_e \rightarrow 0 \), (\ref{we}) implies that \( \Tilde{\omega} \rightarrow R_e^{\T} \omega_d \). Combining this with \( R_e \rightarrow I \) results in \( \Tilde{\omega} \rightarrow \omega_d \), thus solving Problem \ref{innerloop problem}, see  \cite{Alexpaper} for more details.

\section{OUTER-LOOP DYNAMICS AND CONSTRAINTS}
\label{sec:sec4b}
\begin{figure}[!t]
\centering
\includegraphics[width=2in]{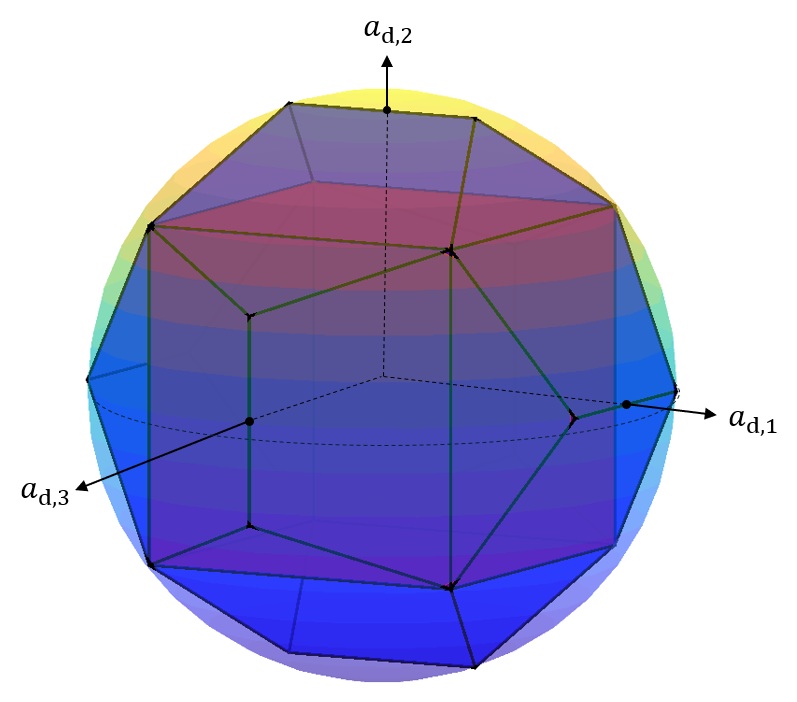}
\caption{Schematic of approximations for the admissible set of \(a_d\): the ideal set as a sphere, approximated either by a dodecahedron (blue) with a circumscribed sphere of radius \(\rho(t)\)
, representing the coupled constraints, or by a cube (red) representing the decoupled constraint region $(|a_{d,i}(t)| \leq \frac{1}{\sqrt{3}}\rho (t),i\in\{1,2,3\}\})$, highlighting its conservative nature.
}
\label{dodecahydron}
\vspace{-0.4cm}
\end{figure}

Since the outer-loop MPC operates in discrete time and holds the control input constant between sampling intervals, its output cannot be directly used to generate the twice differentiable desired acceleration. Therefore, to ensure that $a_d$
remains twice differentiable, the translational error dynamics (\ref{outerloop dynamic}) are extended to a 12th-order linear model (\hspace{-0.001 cm}\cite{izadi2024}) 
\vspace{-0.1 cm}
\begin{subequations}\label{Main outerloop dynamics}
\begin{align}
\Dot{\Tilde{p}}&=\Tilde{v}, \label{position error dynamic} \\
\Dot{\Tilde{v}}&=-D\Tilde{v}+a_d,\label{velocity error dynamic} \\
 \Dot{a}_d&=-\frac{1}{\gamma}(a_d-\eta),
 \label{ad dynamic} \\
\Dot{\eta}&=-\frac{1}{\gamma}(\eta-s), \label{eta dynamic}
\end{align}
\end{subequations}
with $\gamma>0$,  $\Tilde{p}, \Tilde{v}, a_d, \eta \in \mathbb{R}^3$ are the states and $s \in \mathbb{R}^3$ is the input. As detailed in Section \ref{section 4.1}, $a_d$ must lie in the admissible set defined in (\ref{main ad constraint}). In \cite{Alexpaper}, it is shown that a new and smaller set of admissible desired acceleration values, $\mathcal{A}_o(\bar{T}) \subset \mathcal{A}(\bar{R}, \bar{T})$, can be defined as 
\begin{equation}
\mathcal{A}_o(\bar{T})=\{a_d \in \mathbb{R}^3 ~|~  \| a_d \| \leq \rho (t)\},
   \label{TV cons coupled}
\end{equation}
where
\begin{equation}
\label{R(t)}
    \rho (t)=\min(\bar{T}(t)-\delta~,~T_{\mathrm{max}}-\bar{T}(t)).
\end{equation}
The set in (\ref{TV cons coupled}) defines two spheres for the desired acceleration $a_d$ to be in of radius $\bar{T}(t)-\delta$ and $T_{\mathrm{max}}-\bar{T}(t)$, respectively. The approach proposed in \cite{izadi2024} decouples the nonlinear constraint in (\ref{TV cons coupled}), approximating the set (\ref{TV cons coupled}) by the largest cube that fits within the smaller sphere at each time instance. See Fig. \ref{dodecahydron}. Although decoupling the constraints in the $x$, $y$, and $z$ directions simplifies control design and reduces effort by solving three smaller problems, it introduces conservatism due to the conservative nature of the decoupled constraints. Note that in \cite{Alexpaper}, the decoupled constraint is made time-invariant by considering the smallest cube over all time, which introduces additional conservatism.

In our novel approach here, instead of \enquote{decoupling}, we approximate the sphere in (\ref{TV cons coupled}) by a TV dodecahedron with a circumscribed sphere of radius $\rho(t)$, which serves as the admissible set for $a_d$, and results in \enquote{coupled} constraints. This reduces the conservatism in the constraints compared to the approach outlined in \cite{izadi2024}, as shown in Fig. \ref{dodecahydron}. So, the approximated set $\mathcal{A}_{d}(\bar{T})\subset  \mathcal{A}_o(\bar{R},\bar{T})$ is considered where
\begin{align}
    &\mathcal{A}_{d}(\bar{T})= \big\{ a_d \in \mathbb{R}^3\, \big|~ 
     \big|\frac{\sqrt{3}}{\Phi^2} a_{d,1} \pm \frac{\sqrt{3}}{\Phi} a_{d,2}\big| \leq \rho (t),
\big|\frac{\sqrt{3}}{\Phi^2} a_{d,2}  \nonumber \\ & \pm\frac{\sqrt{3}}{\Phi} a_{d,3}\big| \leq \rho (t),~
    \big|\frac{\sqrt{3}}{\Phi^2} a_{d,3} \pm \frac{\sqrt{3}}{\Phi} a_{d,1}\big| \leq \rho (t)
    \big\},
    \label{TV dodecahydron}
\end{align}
with $\Phi=\frac{1+\sqrt{5}}{2}$ and $\rho (t)$ defined in (\ref{R(t)}).

\section{MPC FORMULATION}
\vspace{-0.05cm}\label{sec:sec5}
The MPC law will be formulated in discrete time, thereby requiring the conversion of (\ref{Main outerloop dynamics}) to a discrete-time model. Using exact zero-order hold (ZOH) discretization, with $s(t)=s(t_k),~t\in[t_k,t_{k+1})$, $t_k=kh$, $k \in \mathbb{N}_0$, and $h>0$ representing the sampling period, leads to 
\begin{equation}
    x^+=A_d x+ B_d u,
    \label{DT system dynamics}
\end{equation}
where $x=x(t_k)$, $x=[\Tilde{p}~ \Tilde{v}~ a_d~ \eta]^{\T} \in \mathbb{R}^{12}$ representing the system states, and $u=s(t_k) \in \mathbb{R}^3$ denotes the system input. The system matrices $A_d$ and $B_d$ are presented in \cite{izadi2024}. Note that $A_d$ has nine eigenvalues inside the unit circle and three simple eigenvalues on the unit circle, indicating marginal stability of the system. 
The main continuous-time problem is subject to TV state constraints defined in (\ref{TV dodecahydron}). However, to obtain a discrete-time MPC setup, this constraint must be transformed into specific constraints on the sample times, while ensuring inter-sample constraints satisfaction.
The following MPC problem is formulated for the outer loop:
\vspace{-0.5cm}
\begin{subequations}
\label{MPC problem 2}
 \begin{align}
    \min_{U_k}~&~ J(x_k,U_k) = V(x_{N|k})+\sum_{i=0}^{N-1}(x_{i|k}^{\T}Qx_{i|k}+u_{i|k}^{\T}R ~u_{i|k})\nonumber\\
   \mathrm{s.t.}~&~ x_{0|k}=~x_k,\nonumber\\
    &~x_{i+1|k}=~A_d x_{i|k}+B_d u_{i|k},~i\in \{0,1,..,N-1\},\nonumber\\
    &~u_{i|k} \in  ~\mathscr{U}_{i|k},~~i\in \{0,1,..,N-1\},\label{mpccons-a}\\
     &~a_{d_{i|k}} \in  ~\mathscr{U}_{i|k},~i\in \{0,1,..,N\},\label{mpccons-b}\\
     &~ ~\eta_{i|k} \in  ~\mathscr{U}_{i|k},~~i\in \{0,1,..,N\},\label{mpccons-c}
 \end{align}
 \end{subequations}
where
\vspace{-0.25cm}
\begin{align*}
    \mathscr{U}_{i|k} = \Big\{ (u_1, u_2, u_3)& \in \mathbb{R}^3 \mid \Big| \frac{\sqrt{3}}{\Phi^2} u_m \pm \frac{\sqrt{3}}{\Phi} u_n \Big| \leq \rho_{i|k}, \\ &\forall (m,n) \in \{(1,2), (2,3), (3,1)\} \Big\},
\end{align*}
and 
\vspace{-0.1cm}
\begin{equation}
\hspace*{-2.5cm} 
    \begin{aligned}
    &\rho_{i|k}:=\min_{t \in [t_{k+i}~,~t_{k+i+1}]} \rho (t),
\end{aligned}
\label{R(k)}
\end{equation}
with $\rho (t)$ in (\ref{R(t)}),  \mbox{$U_k=[u_{0|k}, u_{1|k}~\cdots~u_{N-1|k}]^{\T}$} contains the predicted future control inputs, $N \in \mathbb{N}$ is the prediction horizon, $V:\mathbb{R}^{12} \rightarrow \mathbb{R}_{\geq 0}$ represents the terminal cost and is chosen such that $V$ is a control Lyapunov
function, $Q$ and $R$ are positive definite, $x_k$ is the state at discrete-time step $k$, and $x_{i|k}$, $u_{i|k}$ denote the prediction of the state and input at time step $i+k$, respectively, based on information available at time $k$. The following sections cover inter-sample constraint satisfaction, the conversion of state to input constraints, and the UGAS guarantee of the proposed MPC law.
 \subsection{Inter-sample constraints satisfaction}\label{intersample Section }
Effective tracking requires MPC to account for intersample state behavior, especially with long sampling periods where system states may evolve significantly between control actions. Practically, satisfying constraint (\ref{TV dodecahydron}) between sample times ensures safe quadcopter operation within thrust limits at each moment and is critical for formal stability guarantees. The following lemma proves that the MPC problem (\ref{MPC problem 2}) ensures satisfaction of (\ref{TV dodecahydron}) between sample times.
\begin{lemma}

\label{intersample lemma}
    The MPC law resulting from the optimization problem (\ref{MPC problem 2}) guarantees satisfaction of the constraint (\ref{TV dodecahydron}) between intersample intervals. 
\end{lemma}
\vspace{-0.3cm}
\begin{proof}
See Appendix \ref{app lemma intersample proof}.
\end{proof}
\vspace{-0.1cm}
\subsection{State-to-Input Constraint Mapping}
The MPC problem \ref{MPC problem 2} is subject to two state constraints and one input constraint. In this section, a nonlinear mapping from \cite{statetoinput} is used to transform the \textit{state} constraint sets on the states $a_d$ and $\eta$ into equivalent TV \textit{input} constraint sets. This approach anticipates the impact of current control actions on future state constraints, recalculating admissible control actions to prevent constraint violations. By applying the following two lemmas, (\ref{mpccons-b}) and (\ref{mpccons-c}) are converted into equivalent input constraints. Given the system matrices $A_d$ and $B_d$ from \cite{izadi2024} for system (\ref{DT system dynamics}), the discrete-time dynamics of the states $a_d$ and $\eta$ are formulated as
\begin{align}
a_d(k+1)&=\alpha a_d(k)+\beta \eta(k)+[1-\alpha-\beta] u(k), \label{ad dynamics} \\
\eta(k+1)&=\alpha \eta(k)+[1-\alpha]u(k), \label{eta dynamics} 
\end{align}
where $\alpha=e^{-\frac{h}{\gamma}}$,  $\beta=\frac{h}{\gamma}e^{-\frac{h}{\gamma}}$, satisfying $0 < \alpha\leq 1$, $0 < \beta\leq e^{-1}$, and $0< \alpha+\beta\leq 1$.
\begin{lemma}
\label{ad TV lemma}
Consider the MPC problem defined in \eqref{MPC problem 2}. If the TV input constraint
\begin{equation}
\label{R tilde cons}
    \Tilde{\rho}^-_{j_{i|k}} \leq \frac{\sqrt{3}}{\Phi^2} ~u_{m_{i|k}} ~+~(-1)^{j+1} \frac{\sqrt{3}}{\Phi} ~u_{n_{i|k}} \leq \Tilde{\rho}^+_{j_{i|k}},
\end{equation}
holds for all $i \in \{0,1,..,N-1\}$, where 
\begin{align}
\label{R tilda +-}
\Tilde{\rho}&^{\pm}_{j_{i|k}}:=(1-\alpha-\beta)^{-1}
        \big[\pm\rho_{i+1|k}-\alpha\big(\frac{\sqrt{3}}{\Phi^2} a_{d_{m_{i|k}}}+(-1)^{j+1} \nonumber\\
        &\frac{\sqrt{3}}{\Phi}a_{d_{n_{i|k}}}\big)
        -\beta\big(\frac{\sqrt{3}}{\Phi^2} ~\eta_{m_{i|k}} +  (-1)^{j+1}\frac{\sqrt{3}}{\Phi}\eta_{n_{i|k}} \big)    \big],
\end{align}
with $\rho_{i+1|k}$ in (\ref{R(k)}), $j \in \{1,2,..,6\}$ and the indices $m$ and $n$ are determined as follows: if $j \in \{1,2\}$, then  $m=1$ and $n=2$ ; if $j \in \{3,4\}$, then  $m=2$ and $n=3$; and if $j \in \{5,6\}$, then $m=3$ and $n=1$, then it also holds that
\begin{equation*}
    a_{d_{i+1|k}} \in  ~\mathscr{U}_{i+1|k}.
\end{equation*}
\end{lemma}
\vspace{-0.3cm}
\begin{proof}
See Appendix \ref{ad TV lemma proof}.
\end{proof}

\begin{lemma}
\label{eta TV lemma}
Consider the MPC problem defined in \eqref{MPC problem 2}. If the TV input constraint
\vspace{-0.2 cm}
\begin{equation}
\label{R bar cons}
    \bar{\rho}^-_{j_{i|k}} \leq \frac{\sqrt{3}}{\Phi^2} ~u_{m_{i|k}} ~+~ (-1)^{j+1}\frac{\sqrt{3}}{\Phi} ~u_{n_{i|k}} \leq \bar{\rho}^+_{j_{i|k}},
\end{equation}
holds for all $i \in \{0,1,..,N-1\}$, where 
\vspace{-0.1cm}
\begin{align}
\label{R bar +-}
    \bar{\rho}^{\pm}_{j_{i|k}}:=(1-\alpha)^{-1}
        \big[\pm \rho_{i+1|k}-&\alpha\big(\frac{\sqrt{3}}{\Phi^2} ~\eta_{m_{i|k}}+\nonumber\\
        & (-1)^{j+1}\frac{\sqrt{3}}{\Phi}\eta_{n_{i|k}} \big)    \big],
\end{align}
with $\rho_{i+1|k}$ in (\ref{R(k)}), $j \in \{1,2,..,6\}$ and the indices $m$ and $n$ defined as in Lemma \ref{ad TV lemma}, then it also holds that
\vspace{-0.2cm}
\begin{equation*}
    \eta_{i+1|k} \in  ~\mathscr{U}_{i+1|k}.
\end{equation*}
\end{lemma}
\vspace{-0.3cm}
\begin{proof}
See Appendix \ref{eta TV lemma proof}.
\end{proof}
Using Lemmas \ref{ad TV lemma} and \ref{eta TV lemma}, the state constraints (\ref{mpccons-b}) and (\ref{mpccons-c}) are converted into input constraints. This allows the unification of all three constraints (\ref{mpccons-a})–(\ref{mpccons-c}) in MPC problem (\ref{MPC problem 2}) into a single input constraint set as
\vspace{-0.1cm}
\begin{align}
\label{input set irregular dodecahedron}
    \begin{split}
\mathscr{U}'_{i|k}&=\big\{ ((u_x,u_y,u_z) \in \mathbb{R}^3 \big|~\\
     \rho^-_{j_{i|k}} &\leq \frac{\sqrt{3}}{\Phi^2} ~u_{m_{i|k}} ~+~ (-1)^{j+1}\frac{\sqrt{3}}{\Phi} ~u_{n_{i|k}}  \leq \rho^+_{j_{i|k}}\big\}, 
\end{split}
\end{align}
with
\vspace{-0.3cm}
\begin{align}
         \rho^-_{j_{i|k}}&=\max(\Tilde{\rho}^-_{j_{i|k}},\bar{\rho}^-_{j_{i|k}},-\rho_{i|k}),
         \label{R-}\\
         \rho^+_{j_{i|k}}&=\min(\Tilde{\rho}^+_{j_{i|k}},\bar{\rho}^+_{j_{i|k}},+\rho_{i|k}),
         \label{R+}
\end{align}
$j \in \{1,2,..,6\}$, the indices $m$ and $n$ defined as in Lemma \ref{ad TV lemma}, and $\Tilde{\rho}^-_{j_{i|k}}$ and $\Tilde{\rho}^+_{j_{i|k}}$ and $\bar{\rho}^-_{j_{i|k}}$ and $\bar{\rho}^+_{j_{i|k}}$ defined in (\ref{R tilda +-}) and (\ref{R bar +-}). The TV input set $\mathscr{U}'_{i|k}$ can be visualized as an irregular dodecahedron. Theorem \ref{TV theorem} establishes a necessary condition for the feasibility of the MPC problem (\ref{MPC problem 2}) with the input constraint (\ref{input set irregular dodecahedron}), ensuring that the input set is not empty.
\begin{theorem}
\label{TV theorem}
If $h$ and $\gamma$ are appropriately chosen such that 
\begin{equation}
    \rho (k+1) > e^{-\frac{h}{\gamma}} \left(1 + \frac{h}{\gamma}\right) \rho(k), \quad \forall k \in \mathbb{N}_0,  
    \label{h/gam cons}
\end{equation}
holds for a given finite-time trajectory with \(\rho(k) = \rho_{0|k}\) in (\ref{R(k)}), then the unified input constraint in (\ref{input set irregular dodecahedron}) is non-empty, and the MPC controller \eqref{MPC problem 2} with the unified input constraint (\ref{input set irregular dodecahedron}) will be feasible for all $x \in \mathbb{R}^{12}$.
\end{theorem}
\begin{proof}
\vspace{-0.2cm}
See Appendix \ref{TV theorem proof}.
\end{proof}
\subsection{Stability Analysis}
This section proves the UGAS property of the tracking error system in the outer loop. By leveraging the marginal stability of the outer-loop dynamics (\ref{DT system dynamics}), an appropriate cost function can be used to find a Lyapunov function that guarantees the UGAS of the closed-loop system. Following \cite{izadi2024}, we employ the MPC proposed in \cite{neutrallystable} that globally stabilizes marginally stable linear systems under input constraints using a particular terminal cost combining cubic and quadratic state terms. Since the MPC problem (\ref{MPC problem 2}) involves TV input constraints, we derive a smaller time-invariant input set in the following lemma solely for the stability proof. This set serves as the saturation function (\ref{sat function}), enabling the application of Theorem \ref{Thoery small gain} to determine the Lyapunov function.
\begin{lemma}
\label{smallest input set lemma}
Consider the TV input set (\ref{input set irregular dodecahedron}) with $\rho^-_{j_{i|k}}$ and $\rho^+_{j_{i|k}}$ defined in $\eqref{R-}$ and $\eqref{R+}$, then the following time-invariant input set $\mathscr{U}''$, visualized as a regular dodecahedron, lies within the $\mathscr{U}'_{i|k}$ at all sample times $k\in \mathbb{N}_0$:
    \vspace{-0.1cm}
\begin{align}
\label{Time_invarient input set}
    \mathscr{U}''= \Big\{ (u_1, u_2, &u_3) \in \mathbb{R}^3 \mid \big| \frac{\sqrt{3}}{\Phi^2} u_m \pm \frac{\sqrt{3}}{\Phi} u_n \big| \leq \rho^*,\\
    &\forall (m,n) \in \{(1,2), (2,3), (3,1)\} \Big\},\nonumber
\end{align}
    \vspace{-0.1cm}
where $\rho^*$ is a positive constant value obtained by
\begin{equation}
\label{R*}
    \rho^*= \min(\Tilde{\rho}^+_{\mathrm{min}} ,~ \bar{\rho}^+_{\mathrm{min}},~\rho_{\mathrm{min}})~, 
    \vspace{-0.2cm}
\end{equation}
and
\begin{align}
        \Tilde{\rho}^+_{\mathrm{min}}&=(1-\alpha-\beta)^{-1} \min_{k}[\rho(k+1)-(\alpha+\beta)\rho (k)],\nonumber\\
        \bar{\rho}^+_{\mathrm{min}}&=(1-\alpha)^{-1}\min_{k} (\rho(k+1)-\alpha \rho (k)),\nonumber\\
        \rho_{\mathrm{min}}&=\min_{k} \rho (k),
        \label{R star bound}
   \end{align}
$\alpha=e^{-\frac{h}{\gamma}}$,  $\beta=\frac{h}{\gamma}e^{-\frac{h}{\gamma}}$, and $\rho (k)=\rho_{0|k}$ in (\ref{R(k)}).
\end{lemma}
\begin{proof}
See Appendix \ref{smallest input set lemma proof}.
\end{proof}
Theorem \ref{MPC2 stablity theorem} is the main contribution of this section, outlining an MPC strategy with TV coupled input constraints that guarantees UGAS and inter-sample constraint satisfaction.
\begin{theorem}
    \label{MPC2 stablity theorem}
    Consider the closed-loop system \eqref{DT system dynamics}, where the system is marginally stable and the pair $(A_d,B_d)$ is controllable, with the  optimal control sequence $U^*_k=\big(u^*_{0|k}(x), u^*_{1|k}(x),...,u^*_{N|k}(x)\big)$ and MPC control law 
$k_N(x_k)=u^*_{0|k}(x)$ resulting from the optimization problem
\vspace{-0.3cm}
\begin{align}   
\label{stable MPC problem 1_final}
    \min_{U_k}~&~ J(x_k,U_k) = V(x_{N|k})+\sum_{i=0}^{N-1}(x_{i|k}^{\T}Qx_{i|k}+u_{i|k}^{\T}R ~u_{i|k})\nonumber\\
    \mathrm{s.t.}~&~ x_{0|k}=~x_k,\\
    &~x_{i+1|k}=~A_d x_{i|k}+B_d u_{i|k},~i\in \{0,1,..,N-1\},\nonumber\\
    &~u_{i|k} \in  ~\mathscr{U}'_{i|k},~~~~~~~~~~~~~~~~~~i\in \{0,1,..,N-1\},\nonumber
\end{align} 
where $\mathscr{U}'_{i|k}$  defined in (\ref{input set irregular dodecahedron}) and
\begin{equation*}
\hspace*{0cm} 
\begin{aligned}
    & V(x)=\Theta W(x)=\Theta\big[x^{\T} M_q x+\lambda(x^{\T} M_c x)^{3/2}\big],
\end{aligned}
\end{equation*}
with $M_c$ and $M_q$ are as in \eqref{Mc} and \eqref{Mq}, $N \in \mathbb{N}$ is the prediction horizon, $Q$ and $R$ are positive definite matrices, and $\Theta$ is a positive constant such that
\vspace{-0.1cm}
\begin{equation}
    \Theta\geq \lambda_{max}(Q+\kappa^2 A^{\T}_d M_c B_d R B_d^{\T} M_c A_d).
    \label{theta}
\end{equation}
\vspace{-0.1cm}
Furthermore, let $h$ and $\gamma$ be chosen such that
\eqref{h/gam cons} holds for the given trajectory, with $\rho(k)=\rho_{0|k}$ in (\ref{R(k)}). Then for any positive integer $N$, the closed-loop system $x^+=A_dx+B_dk_N(x)$ is uniformly globally asymptotically stable for
\begin{equation*}
\lambda=\frac{2 \kappa L_u \sigma_{max}(A_d^{\T} M_q B_d)}{\sqrt{\lambda_{min}(M_c)}},
\end{equation*}
where $\kappa$ satisfies (\ref{kappa}),  and $L_u$ is chosen such that  $\frac{\rho^*}{\sqrt{3}}  L_u >1,$ with $\rho^*$ in (\ref{R*}) and (\ref{R star bound}).
\end{theorem}
\vspace{-0.3cm}
\begin{proof}
See Appendix \ref{MPC2 stablity theorem proof}.
\end{proof}
\vspace{-0.2cm}
\begin{remark}
  Instead of a dodecahedron, any convex polyhedron with a circumscribed sphere of radius $\rho (t)$ can be used as input set $\mathscr{U}_{i|k}$. The reason for approximating a sphere with a convex polyhedron is to ensure intersample constraint satisfaction. Accordingly, stability guarantees can be obtained using the same approach, with the saturation function defined as the largest cube inscribed in the TV polyhedron. A crucial trade-off arises in choosing the polyhedron: a polyhedron with more facets yield a closer approximation to the sphere and improve performance, but at the cost of higher computational complexity. Notably, the cube-based approximation in \cite{izadi2024} can be viewed as a special case of our proposed approach, which leads to more conservative constraints and, consequently, reduced performance.
  \vspace{-0.1cm}
\end{remark}
\begin{remark}
    Due to the cubic term in the terminal cost, the MPC problem \eqref{stable MPC problem 1_final} is not QP, but remains convex and is solvable with convex optimization solvers. 
\end{remark}
\vspace{-0.1cm}
This section establishes UGAS for the proposed outer-loop MPC strategy, while the inner-loop controller is shown to be ULES and UaGAS in \cite{ErijenNLcontroller}. Moreover, in \cite{izadi2024}, UGAS for the continuous-time system \eqref{Main outerloop dynamics} and UaGAS for the full cascaded system were proven under decoupled constraints and dynamics. these proofs remain valid and directly extend to the proposed MPC with coupled dynamics and constraints.
\vspace{-0.3cm}
\section{NUMERICAL CASE STUDY} 
\label{sec:sec7}
This section compares the performance of the proposed strategy in Section \ref{sec:sec5} with the controllers from \cite{izadi2024} (using decoupled TV constraints), and the controller from \cite{Alexpaper} (using time-invariant decoupled constraints and auxiliary stabilizing constraints) via numerical examples. The results underscore that our new method enhances the performance by reducing conservatism. The outer-loop state prediction uses the dynamics (\ref{DT system dynamics}) to solve the MPC problem. Simulations are conducted in MATLAB over a 25-second trajectory. In the resulting plots, \enquote{\textit{C-MPC}} refers to the coupled MPC strategy from Section \ref{sec:sec5}, \enquote{\textit{D-MPC}} to the decoupled MPC from \cite{izadi2024}, and \enquote{\textit{Baseline MPC}} to the strategy from \cite{Alexpaper}. In all cases, the inner loop uses the nonlinear controller from \cite{ErijenNLcontroller}, as described in Section \ref{inner loop}. Following \cite{Alexpaper} and \cite{izadi2024}, the simulation dynamics in (\ref{dynamics}) use parameters from \cite{Romero2022}: $g = \unit[9.81]{m/s^2}$, $J = \unit[\mathrm{diag}(2.5, 2.1, 4.3)]{gm^2}$, $D =\unit[\mathrm{diag}(0.26, 0.28, 0.42)]{kg/s}$, $\tau_g = [0, 0, 0]^\top$, $A = 0.1I$, $C = 0.5I$, and $T_{\mathrm{max}} = 45.21~\mathrm{m/s}^2$.

The challenging reference trajectory is given by:  
\begin{equation*}
    \Bar{p}(t) = [2\cos(4t)~~2\sin(4t)~~-10+2\sin(2t)]^{\T},~  
    \Bar{\psi}(t) = 0.2t,  
\label{simulationtrajectory}
\end{equation*}  
where \(\Bar{p}(t)\) defines the position reference and \(\Bar{\psi}(t)\) represents the heading reference, indicating the angle between the projection of \(x_B\) onto the \(x_G - y_G\) plane and the \(x_G\) axis. Given that this trajectory is a feasible reference satisfying (\ref{reference}), and leveraging the quadcopter’s differential flatness, it uniquely determines the reference states and inputs. The simulation begins with the same initial conditions as the numerical example in \cite{Alexpaper}. The inner-loop gains are set to \( K_{\omega} = 30J \), \( K_{R} = 70J \), and \( k = [4.5,~5,~5.5] \). The outer-loop parameters are \( h = \unit[0.05]{s} \), \( \gamma = 0.1 \), \( N = 20 \), $Q=\mathrm{diag}(100,1,1,1,100,1,1,1,80,1,1,1)$, and $R=\mathrm{diag}(0.01,0.01,0.1)$. To accelerate the evaluation, CasADi is utilized for solving the MPC problems due to its efficiency in nonlinear optimization and seamless integration with C++, Python, and MATLAB. The MPC function code employs the multiple shooting method to enhance convergence.
\begin{figure}[!t]
\centering
\includegraphics[width=2.4in]{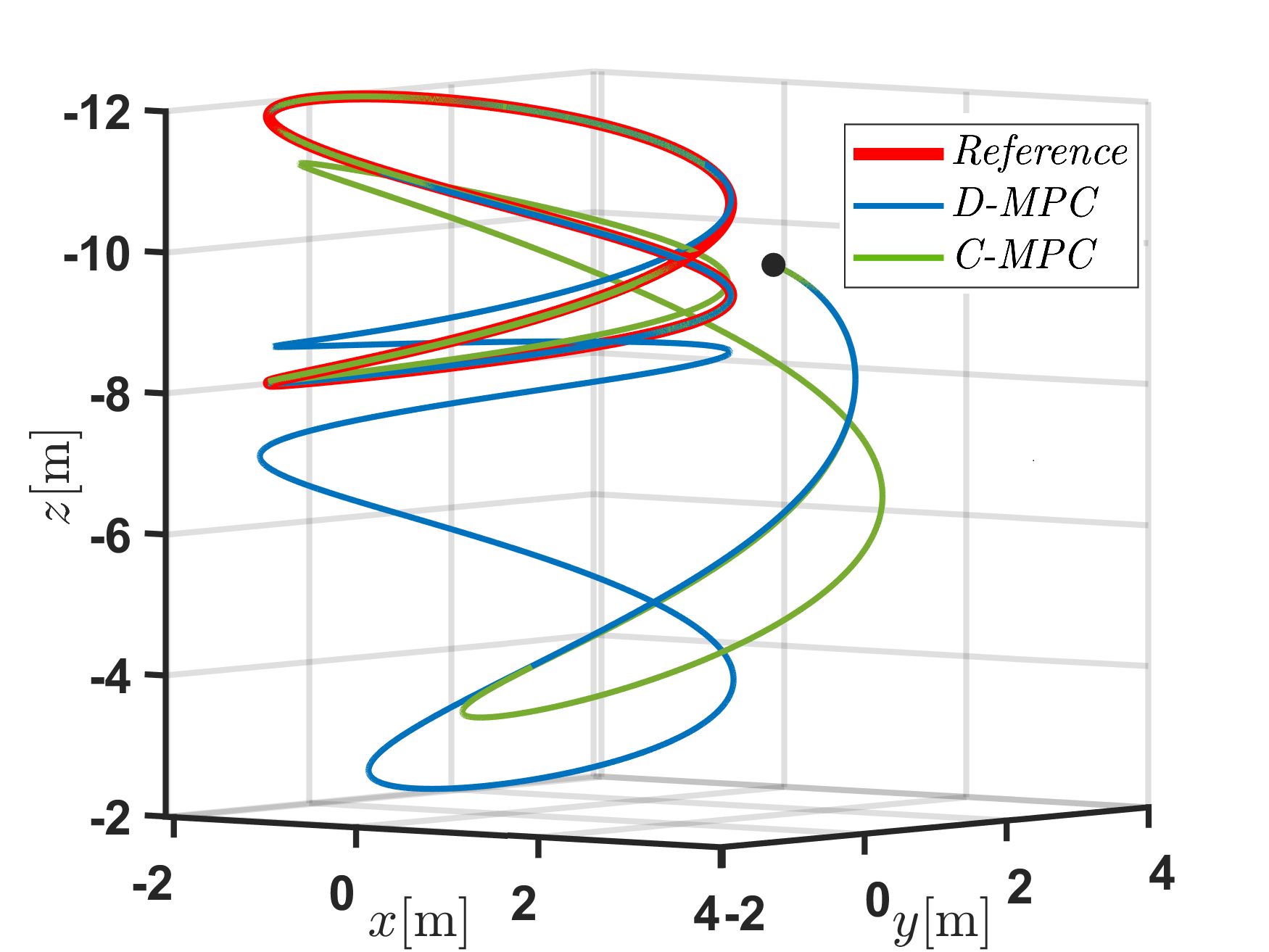}
\caption{3D plot of the reference and actual trajectories, with the actual start marked by a black circle.}
\label{referencetrack sim}

\end{figure}
\begin{figure}[!t]
\centering
\includegraphics[width=2.5in]{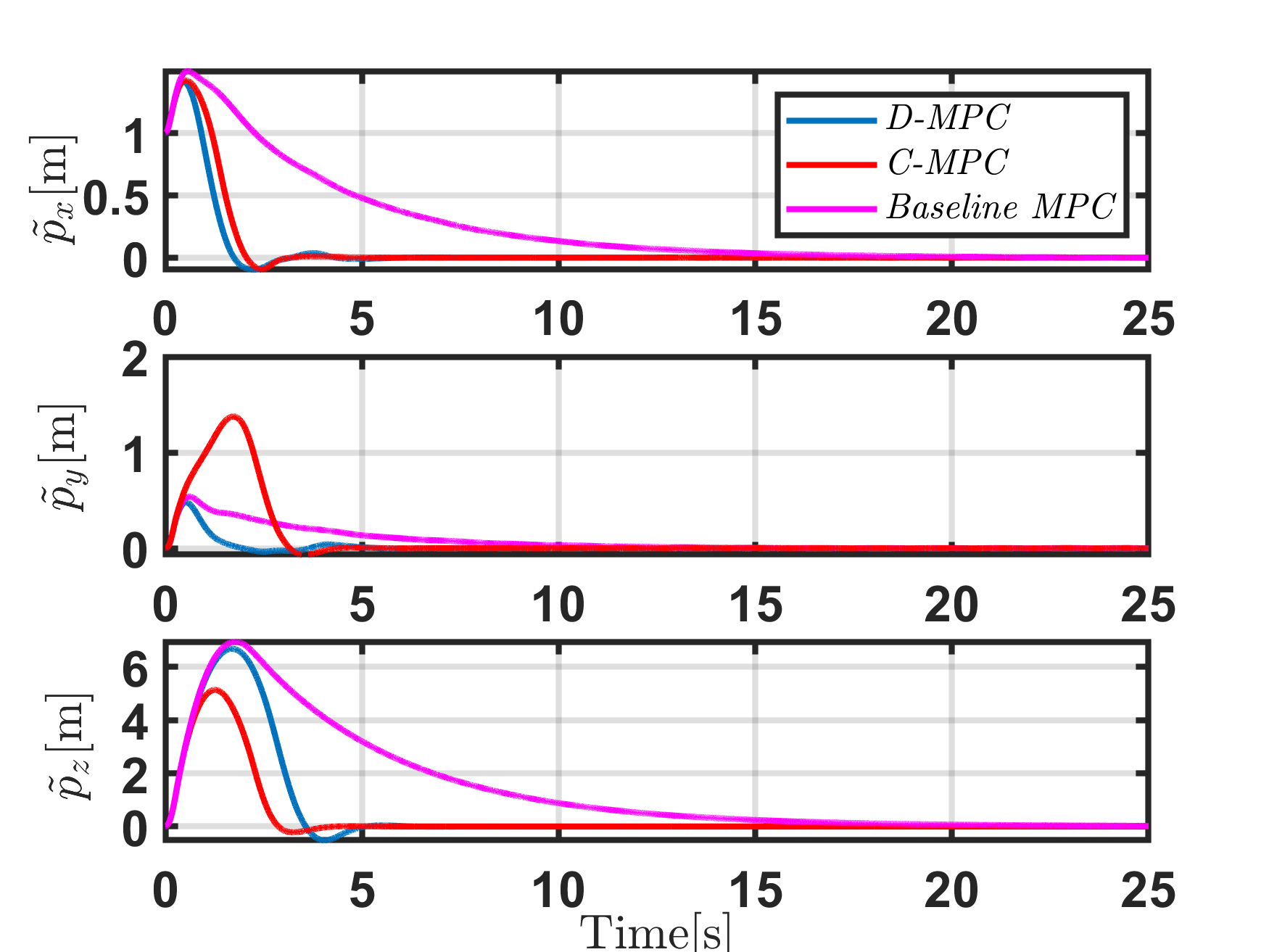}
\caption{Position errors $\Tilde{p}=[\Tilde{p}_x, \Tilde{p}_y, \Tilde{p}_z]$ for the three axes.}
\label{error sim}
\vspace{-0.1cm}
\end{figure}
\begin{figure*}[!t]
\centering
\subfloat[]{\includegraphics[width=2.1in]{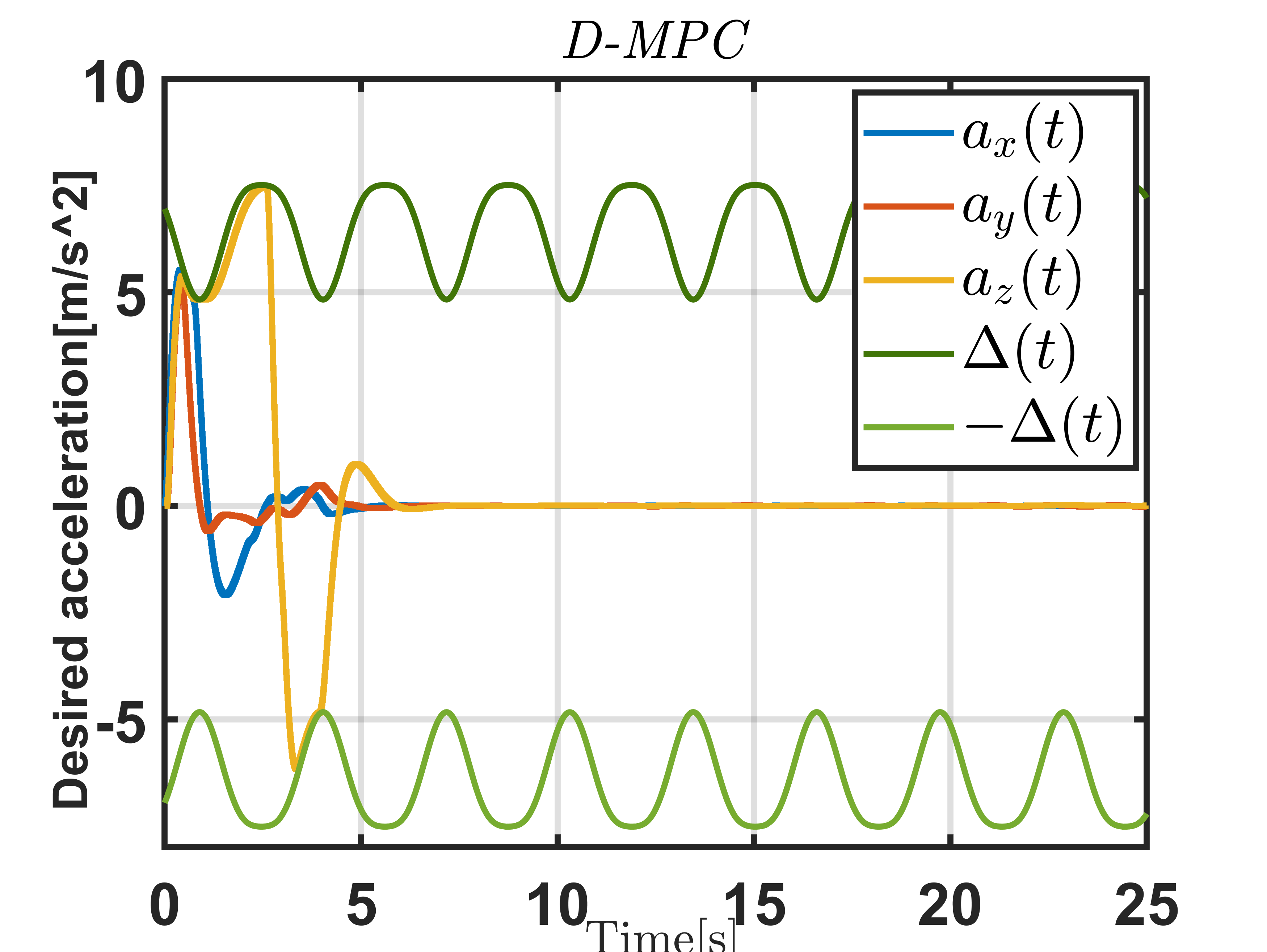}%
\label{MPC 1 ad}}
\hfil
\subfloat[]{\includegraphics[width=2.1in]{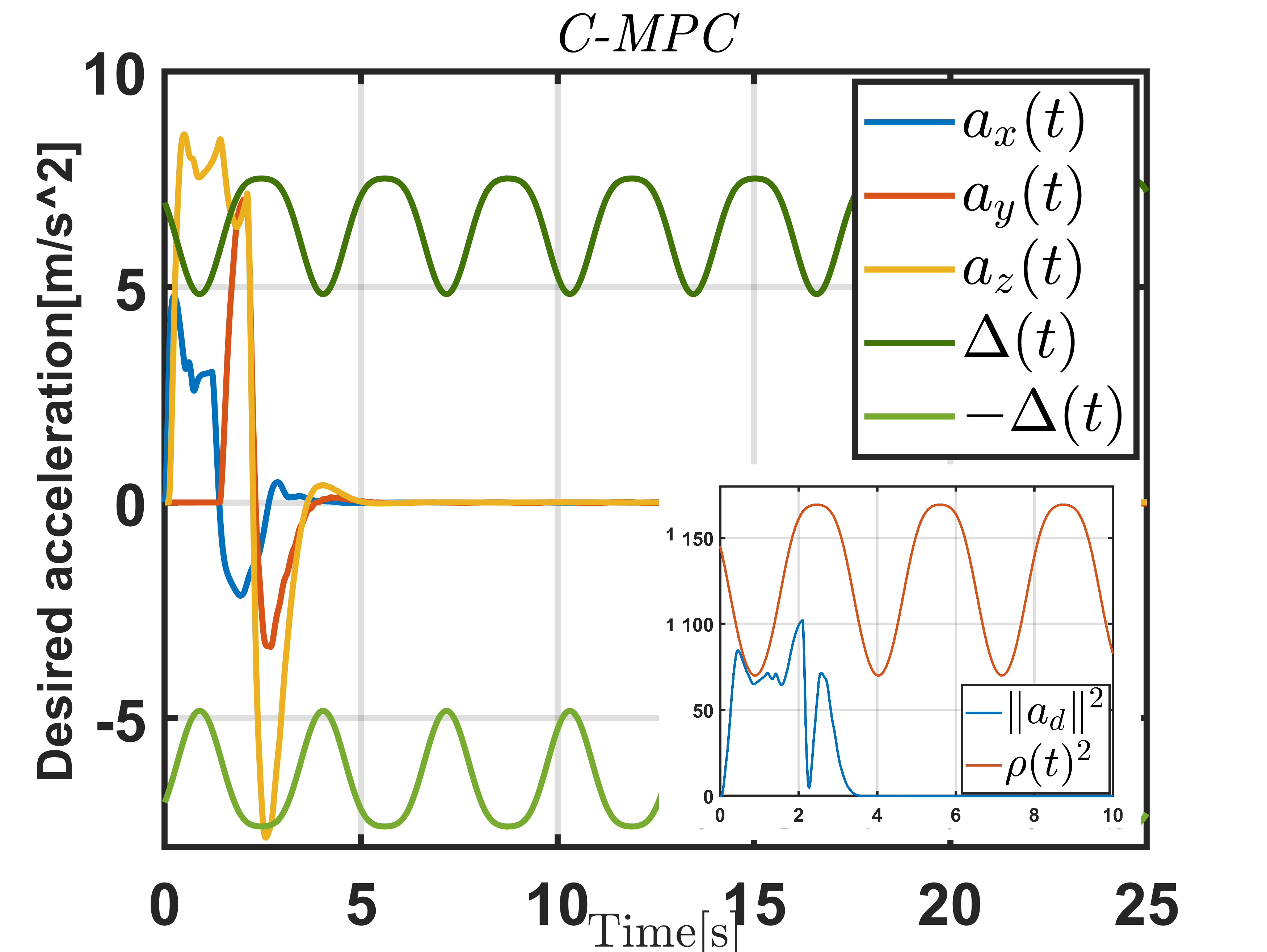}%
\label{MPC 2 ad}}
\hfil
\subfloat[]{\includegraphics[width=2.1in]{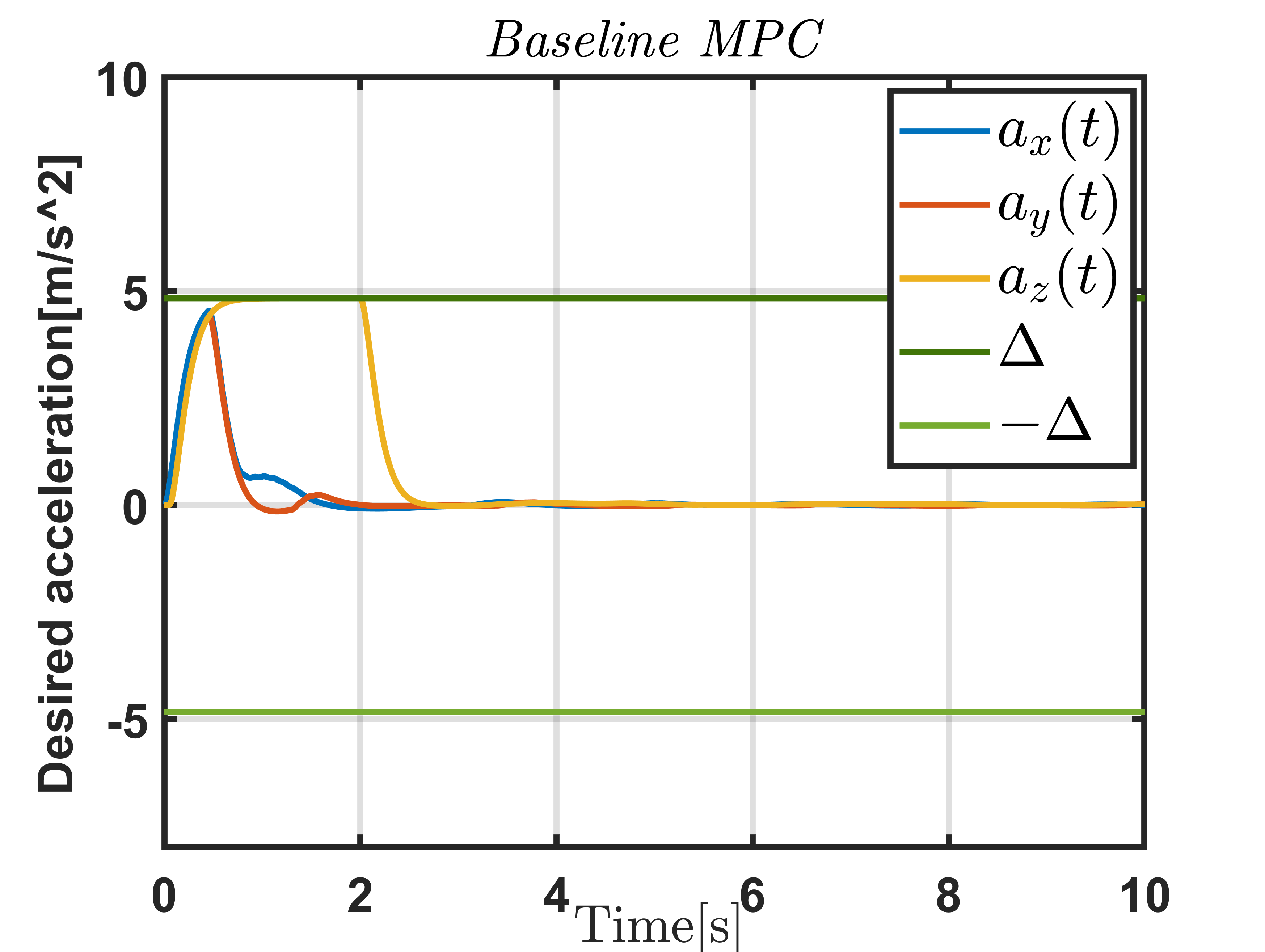}%
\label{MPC 3 ad}}
\caption{Desired accelerations in \(x\), \(y\), and \(z\) directions during the simulation for \textit{C-MPC}, \textit{D-MPC}, and \textit{Baseline MPC}, each imposing different levels of conservatism on the \(a_d\) constraint.  
(a) In \textit{D-MPC}, \(a_{d,i}\) remains within \(\pm \Delta(t) = \pm \frac{\rho(t)}{\sqrt{3}}\) at all times.  
(b) In \textit{C-MPC}, \(a_{d,3}\) occasionally exceeds this range, allowing more flexibility while still satisfying \(\|a_d\|^2 \leq \rho^2(t)\) in \eqref{TV cons coupled}.  
(c) In \textit{Baseline MPC}, \(a_{d,i}\) is restricted to \(\pm \Delta=\pm \min\limits_{t} \frac{\rho(t)}{\sqrt{3}}\), demonstrating its conservative nature.}
\label{ad_sim}
\end{figure*}
\begin{table}[!t]
\caption{Simulation Comparison Results: RMSE [m] and average MPC execution time.  \label{tab:RMSE}}
\centering
\begin{tabular}{|c|c|c|c|}
\hline
 & \textbf{RMSE} & \textbf{Comb. RMSE} & \textbf{Avg. Time}\\
\hline
\hline
\textit{D-MPC} & $[0.26,0.07,1.75]$ & $1.76$ & $3.3$ [ms]\\
\hline
\textit{C-MPC} & $[0.30,0.36,1.12]$ & $1.22$ & $27.3$ [ms]\\
\hline
\textit{Baseline MPC} & $[0.48,0.14,2.43]$ & $2.48$ & $0.25$ [ms]\\
\hline
\end{tabular}
\vspace{-0.6cm}
\end{table}

The reference and resulting position trajectories are shown in Fig. \ref{referencetrack sim}. It is evident that \textit{C-MPC} achieves faster convergence than both \textit{D-MPC} and \textit{Baseline MPC}, thanks to its less conservative constraints. Fig. \ref{error sim} presents the position errors, demonstrating that \textit{C-MPC} outperforms both \textit{D-MPC} and \textit{Baseline MPC}. This underscores the effectiveness of the proposed strategies in reducing conservatism and improving overall performance. Table \ref{tab:RMSE} presents the Root Mean Square Error (RMSE) and average MPC execution time, as measured on a laptop with an Intel\textsuperscript{\textregistered} Core\texttrademark~i7-10750H CPU @ 2.60\,GHz, 16\,GB RAM, and MATLAB R2022b. \textit{C-MPC} achieves the lowest combined RMSE but at the expense of increased computational time. \textit{Baseline MPC} achieves the shortest execution time, as it is a QP problem. However the execution time in all cases is significantly shorter than the MPC sampling period $h=50$ ms. Users face a clear trade-off: \textit{C-MPC} achieves higher performance through less conservative constraints but requires greater computational power, whereas \textit{D-MPC} is feasible on less powerful hardware at the cost of lower performance. Fig. \ref{ad_sim} shows the desired acceleration in the $x$, $y$, and $z$ directions, with the corresponding constraints on $a_d$ satisfied in all cases. \textit{Baseline MPC} applies the most conservative constraint on $a_d$, while \textit{C-MPC} imposes the least conservative one.

\section{CONCLUSIONS AND FUTURE WORK} \label{sec:sec8}
\vspace{-0.1 cm
}This paper has presented an inner-outer loop control structure for quadcopter trajectory tracking with a formal closed-loop tracking guarantee. A novel MPC-based controller has been proposed for the outer loop, effectively handling TV coupled constraints arising from the quadcopter's limited thrust capabilities. A numerical case study has validated the proposed control scheme, showing improved performance due to a less conservative design compared to previous MPC proposals with stability guarantees (\hspace{-0.001cm}\cite{Alexpaper} and \cite{izadi2024}).\\
Future work will further relax the $a_d$ constraints by using the original constraint (\ref{main ad constraint}) instead of the more conservative one in (\ref{TV cons coupled}). In addition, robustness analysis against unmodeled dynamics and disturbances, together with real-world experiments, would help validate the full cascade scheme under practical conditions.

\appendices
\vspace{-0.15 cm}
\section{Proof of lemma \ref{intersample lemma} } \label{app lemma intersample proof}
 Given $A_d$ and $B_d$  presented in \cite{izadi2024}, the evolution of $a_d$ between sampling instants in (\ref{Main outerloop dynamics}) is formulated as
 \vspace{-0.25cm}
    \begin{multline}
        a_{d,i}(t)=\alpha(t-t_k) a_{d,i}(t_k)+\beta(t-t_k) \eta_i(t_k)+\\
        [1-\alpha(t-t_k)-\beta(t-t_k)] u_i(t_k),
        \label{a(t) coupled}
  \end{multline} 
for $t \in [t_k,t_{k+1})$, where $i \in \{1,2,3\}$,  $\alpha(t)=e^{-\frac{t}{\gamma}}$, $\beta(t)=\frac{t}{\gamma}e^{-\frac{t}{\gamma}}$, $0 < \alpha(t)\leq 1$, $0 < \beta(t)\leq e^{-1}$ and $0< \alpha(t)+\beta(t)\leq 1$. Consider the first constraints of (\ref{mpccons-a}-\ref{mpccons-c}) for $i=0$:
$
       -\rho_{0|k} \leq \frac{\sqrt{3}}{\Phi^2} a_{d,1}(k)+\frac{\sqrt{3}}{\Phi} a_{d,2}(k) \leq \rho_{0|k},
    -\rho_{0|k} \leq \frac{\sqrt{3}}{\Phi^2} ~\eta_1(k) +\frac{\sqrt{3}}{\Phi} ~\eta_2(k) \leq \rho_{0|k},
    -\rho_{0|k} \leq \frac{\sqrt{3}}{\Phi^2} ~u_1(k) + \frac{\sqrt{3}}{\Phi} ~u_2(k) \leq \rho_{0|k}.
$
Then for all $t \in [t_k,t_{k+1})$
\begin{align*}
&\frac{\sqrt{3}}{\Phi^2} a_{d,1}(t) + \frac{\sqrt{3}}{\Phi} a_{d,2}(t)=
    \alpha(t-t_k)\Bigr[\frac{\sqrt{3}}{\Phi^2} a_{d,1}(t_k) + \\&\frac{\sqrt{3}}{\Phi} a_{d,2}(t_k)\Bigr]+
\beta(t-t_k)\Bigr[\frac{\sqrt{3}}{\Phi^2} \eta_1(t_k) + \frac{\sqrt{3}}{\Phi} \eta_2(t_k)\Bigr]+ \\
&(1-\alpha(t-t_k)-\beta(t-t_k))\Bigr[\frac{\sqrt{3}}{\Phi^2} u_1(t_k) + \frac{\sqrt{3}}{\Phi} u_2(t_k)\Bigr],
\vspace{-0.2 cm}
\end{align*}
which is the convex combination of $[\frac{\sqrt{3}}{\Phi^2} a_{d,1}(t_k) + \frac{\sqrt{3}}{\Phi} a_{d,2}(t_k)]$, $-[\frac{\sqrt{3}}{\Phi^2} \eta_1(t_k) + \frac{\sqrt{3}}{\Phi} \eta_2(t_k)]$ and $[\frac{\sqrt{3}}{\Phi^2} u_1(t_k) + \frac{\sqrt{3}}{\Phi} u_2(t_k)]$.
Therefore from (\ref{mpccons-a}-\ref{mpccons-c}), it follows that $|\frac{\sqrt{3}}{\Phi^2} a_{d,1}(t) + \frac{\sqrt{3}}{\Phi} a_{d,2}(t)| \leq \rho_{0|k}$ for all  $t \in [t_k,t_{k+1})$. Because $\rho_{0|k}\leq \rho (t)$ for all  $t \in [t_k,t_{k+1}]$ by definition, we can conclude that $|\frac{\sqrt{3}}{\Phi^2} a_{d,1}(t) + \frac{\sqrt{3}}{\Phi} a_{d,2}(t)| \leq \rho (t)$ for all $t$ within this intervals. Inter-sample satisfaction of the remaining five constraints in $\mathscr{U}_{0|k}$ follows the same approach, ensuring $a_d(t) \in \mathcal{A}d(\bar{T})$ for all $t \in [t_k, t{k+1})$.

\section{Proof of lemma \ref{ad TV lemma}} \label{ad TV lemma proof}

Consider the constraints in (\ref{R tilde cons}) for $j=1:
(1-\alpha-\beta)^{-1}[-\rho_{i+1|k}-\alpha(\frac{\sqrt{3}}{\Phi^2} ~a_{d_{1_{i|k}}} + \frac{\sqrt{3}}{\Phi}a_{d_{2_{i|k}}})-
        \beta(\frac{\sqrt{3}}{\Phi^2} ~\eta_{1_{i|k}} + \frac{\sqrt{3}}{\Phi}\eta_{2_{i|k}})]
        \leq \frac{\sqrt{3}}{\Phi^2} ~u_{1_{i|k}} ~+~ \frac{\sqrt{3}}{\Phi} ~u_{2_{i|k}} \leq
        (1-\alpha-\beta)^{-1}[+\rho_{i+1|k}-\alpha(\frac{\sqrt{3}}{\Phi^2} ~a_{d_{1_{i|k}}} + \frac{\sqrt{3}}{\Phi}a_{d_{2_{i|k}}})-
        \beta(\frac{\sqrt{3}}{\Phi^2} ~\eta_{1_{i|k}} + \frac{\sqrt{3}}{\Phi}\eta_{2_{i|k}} )]$. Multiplying $(1-\alpha-\beta)$, a non-negative value, to the above constraint, and then adding $\alpha(\frac{\sqrt{3}}{\Phi^2} ~a_{d_{1_{i|k}}} + \frac{\sqrt{3}}{\Phi}a_{d_{2_{i|k}}})
        +\beta(\frac{\sqrt{3}}{\Phi^2} ~\eta_{1_{i|k}} + \frac{\sqrt{3}}{\Phi}\eta_{2_{i|k}})$ to all three terms of the resulting inequality, yields:
\begin{align*}
    -\rho_{i+1|k}\leq \frac{\sqrt{3}}{\Phi^2} \Big[\alpha a_{d_{1_{i|k}}}+\beta\eta_{1_{i|k}}+(1-\alpha-\beta)u_{1_{i|k}}\Big]+&\\
    \frac{\sqrt{3}}{\Phi}\Big[\alpha a_{d_{2_{i|k}}}+\beta\eta_{2_{i|k}}+(1-\alpha-\beta)u_{2_{i|k}}\Big]\leq \rho_{i+1|k}.&
\end{align*}
Using dynamics of $a_d$ in \eqref{ad dynamics}, we can conclude:
$-\rho_{i+1|k}\leq \frac{\sqrt{3}}{\Phi^2} a_{d_{1_{i+1|k}}}+~
    \frac{\sqrt{3}}{\Phi}a_{d_{2_{i+1|k}}}\leq \rho_{i+1|k}.$
Using the remaining five constraints of (\ref{R tilde cons}) for $j=2,…,6$ and applying the same approach, the following inequalities are derived
\begin{equation*}
    -\rho_{i+1|k}\leq \frac{\sqrt{3}}{\Phi^2} a_{d_{m_{i+1|k}}}+~
    (-1)^{j+1}\frac{\sqrt{3}}{\Phi}a_{d_{n_{i+1|k}}}\leq \rho_{i+1|k},
\end{equation*}
which shows that
   $ a_{d_{i+1|k}} \in  ~\mathscr{U}_{i+1|k}.$
   
\section{Proof of lemma \ref{eta TV lemma}} \label{eta TV lemma proof}
\vspace{-0.1cm}
Consider the constraints in (\ref{R bar cons}) for $j=1: (1-\alpha)^{-1}[-\rho_{i+1|k}-\alpha(\frac{\sqrt{3}}{\Phi^2} ~\eta_{1_{i|k}} + \frac{\sqrt{3}}{\Phi}\eta_{2_{i|k}} )    ]
        \leq
        \frac{\sqrt{3}}{\Phi^2} ~u_{1_{i|k}} ~+~ \frac{\sqrt{3}}{\Phi} ~u_{2_{i|k}} \leq
        (1-\alpha)^{-1}[\rho_{i+1|k}-
        \alpha(\frac{\sqrt{3}}{\Phi^2} ~\eta_{1_{i|k}} + \frac{\sqrt{3}}{\Phi}\eta_{2_{i|k}} )].$ Multiplying $(1-\alpha)$, a non-negative value, to the above constraint, and then adding $\alpha(\frac{\sqrt{3}}{\Phi^2} ~\eta_{1_{i|k}} + \frac{\sqrt{3}}{\Phi}\eta_{2_{i|k}} )$ to all three terms of the resulting inequality, and
applying the same reasoning as in \ref{ad TV lemma proof} yields $\eta_{i+1|k} \in  ~\mathscr{U}_{i+1|k}$.

\section{Proof of theorem \ref{TV theorem}} \label{TV theorem proof}
If the unified input constraint (\ref{input set irregular dodecahedron}) holds, then
\begin{subequations}
\begin{align}
        \Tilde{\rho}^-_{j_{i|k}} \leq& \frac{\sqrt{3}}{\Phi^2} ~u_{m_{i|k}} ~+~ (-1)^{j+1}\frac{\sqrt{3}}{\Phi} ~u_{n_{i|k}} \leq \Tilde{\rho}^+_{j_{i|k}},
        \label{first input const}\\
        \bar{\rho}^-_{j_{i|k}} \leq& \frac{\sqrt{3}}{\Phi^2} ~u_{m_{i|k}} ~+~(-1)^{j+1} \frac{\sqrt{3}}{\Phi} ~u_{n_{i|k}} \leq \bar{\rho}^+_{j_{i|k}},
        \label{second input const}\\
       - \rho_{i|k} \leq& \frac{\sqrt{3}}{\Phi^2} ~u_{m_{i|k}} ~+~(-1)^{j+1} \frac{\sqrt{3}}{\Phi} ~u_{n_{i|k}} \leq \rho_{i|k}.
        \label{third input const}
   \end{align}
\end{subequations}
By meeting (\ref{first input const}) and (\ref{second input const}) and applying Lemma \ref{ad TV lemma} and \ref{eta TV lemma}, we obtain: 
       $ a_{d_{i|k}} \in ~\mathscr{U}_{i|k},~
        \eta_{i|k} \in ~\mathscr{U}_{i|k},~\forall i\in \{1,..,N\}.$      
So, by initializing the controller with $a_{d_{0|0}} \in \mathscr{U}_{0|0}$ and $\eta_{0|0} \in \mathscr{U}_{0|0}$, it follows that $a_{d_{i|k}} \in \mathscr{U}_{i|k}$ and $\eta_{i|k} \in \mathscr{U}_{i|k}$ for all $i \in \{0,1,...,N\}$.
The intersection of the three input constraint sets (\ref{first input const}-\ref{third input const}) is guaranteed to be non-empty if the lower bounds $\Tilde{\rho}^-_{j_{i|k}}$ and $\bar{\rho}^-_{j_{i|k}}$ are negative, and the upper bounds $\Tilde{\rho}^+_{j_{i|k}}$ and  $\bar{\rho}^+_{j_{i|k}}$ are positive (knowing that $\rho_{i|k}$ is always positive). It can be concluded from $a_{d_{i|k}} \in ~\mathscr{U}_{i|k}$ and
        $\eta_{i|k} \in ~\mathscr{U}_{i|k}$ that 
$-\rho_{i|k} \leq
\frac{\sqrt{3}}{\Phi^2} a_{d_{m_{i|k}}}+ (-1)^{j+1}\frac{\sqrt{3}}{\Phi}a_{d_{n_{i|k}}} \leq \rho_{i|k} $ and 
$-\rho_{i|k} \leq
\frac{\sqrt{3}}{\Phi^2} \eta_{{m_{i|k}}}+ (-1)^{j+1}\frac{\sqrt{3}}{\Phi}\eta_{{n_{i|k}}} \leq \rho_{i|k} $ for all $j \in \{1,2,..,6\}$. It follows that
\begin{align}
        -&(\alpha+\beta)\rho_{i|k} \leq 
        \alpha \Big[\frac{\sqrt{3}}{\Phi^2} a_{d_{m_{i|k}}} + (-1)^{j+1}\frac{\sqrt{3}}{\Phi} a_{d_{n_{i|k}}}\Big]+ \nonumber\\
        &~~\beta \Big[ \frac{\sqrt{3}}{\Phi^2} \eta_{m_{i|k}} + (-1)^{j+1} \frac{\sqrt{3}}{\Phi} \eta_{n_{i|k}}\Big]\leq(\alpha+\beta)\rho_{i|k},
        \label{temp1}\\
        -&\alpha \rho_{i|k}\leq 
        \alpha \Big[\frac{\sqrt{3}}{\Phi^2} \eta_{m_{i|k}} + (-1)^{j+1} \frac{\sqrt{3}}{\Phi} \eta_{n_{i|k}}] 
        \leq \alpha \rho_{i|k}.
        \label{temp2}
    \end{align}
   If $h$ and $\gamma$ are chosen such that $\rho_{i+1|k} > (\alpha+\beta) \rho_{i|k}$ for all $k \in \mathbb{N}_0$, then it can be inferred from (\ref{temp1}) and (\ref{temp2}) that
   \begin{align*}
        -\rho_{i+1|k} \leq 
        \alpha &\Big[\frac{\sqrt{3}}{\Phi^2} a_{d_{m_{i|k}}} + (-1)^{j+1}\frac{\sqrt{3}}{\Phi} a_{d_{n_{i|k}}}\Big]+ \nonumber\\
        \beta &\Big[ \frac{\sqrt{3}}{\Phi^2} \eta_{m_{i|k}} + (-1)^{j+1}\frac{\sqrt{3}}{\Phi} \eta_{n_{i|k}}\Big]\leq\rho_{i+1|k},\\
        - \rho_{i+1|k}\leq 
        \alpha &\Big[\frac{\sqrt{3}}{\Phi^2} \eta_{m_{i|k}} +(-1)^{j+1}\frac{\sqrt{3}}{\Phi} \eta_{n_{i|k}}\Big] 
        \leq  \rho_{i+1|k}.
    \end{align*}
From the inequalities above and using $\Tilde{\rho}^{\pm}_{j_{i|k}}$ and $\Bar{\rho}^{\pm}_{j_{i|k}}$  in \eqref{R tilda +-} and (\ref{R bar +-}), it is evident that $\Tilde{\rho}^-_{j_{i|k}}$ and $\bar{\rho}^-_{j_{i|k}}$ are negative, and 
$\Tilde{\rho}^+_{j_{i|k}}$ and $\bar{\rho}^+_{j_{i|k}}$ are positive.
Therefore, the condition $\rho_{i+1|k} > (\alpha+\beta) \rho_{i|k}$, is sufficient to ensure (\ref{input set irregular dodecahedron}) is non-empty.
\section{Proof of lemma \ref{smallest input set lemma}} \label{smallest input set lemma proof}
Using \eqref{R-} and \eqref{R+}, it can be demonstrated that:
\begin{align*}
       \max\limits_{k} \rho^-_{j_{i|k}}=&~ 
       \max~(\max\limits_{k} -\rho_{i|k}, \max\limits_{k} \Tilde{\rho}^-_{j_{i|k}}, \max\limits_{k} \bar{\rho}^-_{j_{i|k}}),\\
        \min\limits_{k} \rho^+_{j_{i|k}}=&~ \min~(\min\limits_{k} \rho_{i|k},  \min\limits_{k} \Tilde{\rho}^+_{j_{i|k}}, \min\limits_{k} \bar{\rho}^+ _{j_{i|k}}),
\end{align*} 
for all $j \in \{1,2,..,6\}$. According to (\ref{R tilda +-}) and (\ref{R bar +-}), using \eqref{temp1} and \eqref{temp2}, and knowing that  $a_{d_{i|k}} \in ~\mathscr{U}_{i|k}$ and $\eta_{i|k} \in ~\mathscr{U}_{i|k}$, for all $i\in \{0,1,..,N-1\}$, it can be shown that:
\begin{align*}
    \begin{split}
       \min\limits_{k} \Tilde{\rho}^+_{j_{i|k}} \geq (1-\alpha-\beta)^{-1} \min_{k}[\rho(k+1)-(\alpha+\beta)\rho (k)]\\
       =\Tilde{\rho}^+_{\mathrm{min}}
    \end{split}\\
    \begin{split}
    \min\limits_{k} \bar{\rho}^+ _{j_{i|k}} \geq (1-\alpha)^{-1}\min_{k} (\rho(k+1)-\alpha \rho (k))=\bar{\rho}^+_{\mathrm{min}}
    \end{split}\\
    \begin{split}
       \max\limits_{k} \Tilde{\rho}^-_{j_{i|k}} \leq (1-\alpha-\beta)^{-1} \min_{k}[-\rho(k+1)+(\alpha+\beta)\rho (k)]\\
       =-\Tilde{\rho}^+_{\mathrm{min}}
    \end{split}\\
    \begin{split}
    \max\limits_{k} \bar{\rho}^- _{j_{i|k}} \leq (1-\alpha)^{-1}\min_{k} (-\rho(k+1)+\alpha \rho (k))=-\bar{\rho}^-_{\mathrm{min}}
    \end{split}
\end{align*} 
and therefore
\begin{align*}
    \begin{split}
       \max\limits_{k} \rho^-_{j_{i|k}}\leq&~ 
       \max~(\max\limits_{k} -\rho_{i|k}, -\Tilde{\rho}^+_{\mathrm{min}}, -\bar{\rho}^-_{\mathrm{min}})=-\rho^*,
    \end{split}\\
    \begin{split}
        \min\limits_{k} \rho^+_{j_{i|k}}\geq& \min~(\min\limits_{k} \rho_{i|k},  \Tilde{\rho}^+_{\mathrm{min}}, \bar{\rho}^-_{\mathrm{min}})=\rho^*,
    \end{split}
\end{align*} 
which shows that dodecahedron $\mathscr{U}''$ with circumscribed sphere of radius $\rho^*$  lies within the $\mathscr{U}'_{i|k}$, i.e., $\mathscr{U}'' \subset  \mathscr{U}'_{i|k}$.
\section{Proof of theorem \ref{MPC2 stablity theorem}} \label{MPC2 stablity theorem proof}
Due to the definition of the saturation function in (\ref{sat function}), the saturation region must be defined as a cube in $\mathbb{R}^3$.
Consider the saturation function (\ref{sat function}) with $u_{\mathrm{max}}=\frac{\rho^*}{\sqrt{3}}$, which can be visualized as the largest cube that fits within the dodecahedron-shaped input set $\mathscr{U}''$. According to Theorem \ref{Thoery small gain}, the local control law $ k_f:\mathbb{R}^{12}  \rightarrow \mathbb{R}$, defined as $k_f(x)=\mathrm{sat}(Kx)$ with
the small gain control $K$ in (\ref{smallgain}), ensures UGAS of the closed-loop system, 
with the global Lyapunov function defined in \eqref{LyapFunc}. 
Knowing that $k_f(x)=\mathrm{sat}(Kx)$, we can infer that 
$k_f(x) \in \mathscr{U}''$ with $\mathscr{U}''$ defined in (\ref{Time_invarient input set}).
By applying Lemma \ref{smallest input set lemma}, it can be inferred that $\mathscr{U}'' \subset \mathscr{U}'_{i|k}$. Demonstrating that $k_f \in \mathscr{U}'_{i|k}$, the assumptions of Theorem 1 and Theorem 3 of \cite{neutrallystable} remain valid for a saturation function with  $u_{\mathrm{max}}=\frac{\rho^*}{\sqrt{3}}$ and $\Theta$ specified in \eqref{theta}. Therefore, the MPC problem (\ref{MPC2 stablity theorem}) is UGAS if $L_u$ is chosen such that  $ \frac{\rho^*}{\sqrt{3}}  L_u >1$.


\bibliographystyle{IEEEtran}
\bibliography{IEEEabrv}

\end{document}